\documentclass[twocolumn, prb, showpacs]{revtex4}
\usepackage{graphicx}
\usepackage{dcolumn}
\usepackage{bm}
\usepackage{amsmath}

\begin{document}

\title{Phase diagram of frustrated mixed-spin ladders in the strong-coupling limit}
\author{Shu Chen, Li Wang and Yupeng Wang}
\address{Beijing National Laboratory for Condensed
Matter Physics, Institute of Physics, Chinese Academy of Sciences,
Beijing 100080, P. R. China}

\begin{abstract}
We study the ground-state properties of frustrated Heisenberg ferrimagnetic
ladders with antiferromagnetic exchange interactions and two types of
alternating sublattice spins. In the limit of strong rung couplings, we show
that the mixed spin-1/2 and spin-1 ladders can be systematically mapped onto
a spin-1/2 Heisenberg model with additional next-nearest-neighbor exchanges.
The system is either in a ferrimagnetic state or in a critical spin-liquid
state depending on the competition between the spin exchanges along the legs
and the diagonal exchanges.
\end{abstract}
\pacs{75.10.Jm}
\date{\today}
\maketitle


\section{Introduction}

In the past years, there has been increasing theoretical interest in the
quantum ferrimagnetic systems since the experimental realization of
bimetallic quasi-one-dimensional (quasi-1D) magnets \cite
{Kahn,Hagiwara,Verdaguer}. Pioneering experiments on synthesizing the
bimetallic chain compounds with each unit cell containing two spins were
carried out successfully by Kahn et. al \cite{Kahn}. Typical compounds
include two families of ferrimagnetic chains described by $%
ACu(pba)(H_2O)_{3}.n H_2O$ and $ACu(pbaOH)(H_2O)_3.nH_2O$, where $pba$%
=1,3-propylenebis(Oxamato), $pbaOH$=2-hydroxo-1,3-propylenebis(Oxamato) and
A=Ni, Fe, Co, and Mn \cite{Hagiwara,Verdaguer}. These materials are quasi-1D
bimetallic molecular magnets containing two different transition-metal ions
per unit cell alternatingly distributed on a chain. Most of these materials
are described by Heisenberg mixed-spin models with antiferromagnetic
interactions. This has stimulated theoretical studies on the mixed-spin
systems. A number of recent studies have been focused on ground-state
properties of mixed-spin chain\cite{Brehmer,Pati,Kolezhuk,Tian,Wu} and
ladders\cite{Kolezhuk2,Langari2000,Langari,Trumper,Aristov,Ivanov,Ivanov04}
composed of two kinds of spins with half-integer and integer spins
alternatively. It is well-known that a half-integer antiferromagnetic spin
chain has gapless excitation spectrum and the integer ones show the Haldane
gap in their low energy spectrum, however the 1D ferrimagnets behave
differently and exhibit intriguing quantum spin phases and thermodynamic
properties. One of the intriguing features of the mixed-spin chain is that
the ground state has a long-range ferrimagnetic order.
\begin{figure}[tbp]
\includegraphics[width=3.4in]{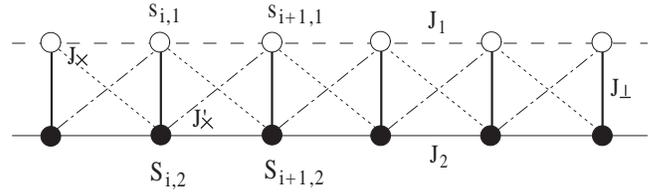}
\caption{Schematical representation of the mixed-spin ladder with diagonal
exchanges. The open and close circles represent spins $s_{_{i,1}}=1/2$ and $%
S_{i,2}=1$ respectively. }
\label{fig2}
\end{figure}

While the physics of the mixed-spin chain is well understood, the spin
ladders are still under intensive investigation since they exhibit rich
phase diagrams. So far, a large class of mixed-spin two-leg ladders are
investigated, but most of studies concern the physics of unfrustrated
mixed-spin ladders. For the unfrustrated bipartite ladder with alternating
half-integer and integer spins on neighboring sites, the ground state is
shown to exhibit long-range order \cite{Tian}. In general, for the uniform
spin ladders, frustration reduces the antiferromagnetic correlations and may
produce various exotic quantum ground states such as the dimerized state\cite
{Chen,Chen2,Lecheminant}. In comparison with the uniform systems, less
attention is paid to the role of frustration in the mixed spin systems and
it is quite interesting to explore how the ferrimagnetic order is affected
by the frustration \cite{Ivanov,Ivanov04}. Recently, the effect of magnetic
frustrations due to diagonal exchange bonds in a mixed-spin ladder was
studied by using the approach of exact numerical diagonalization \cite
{Ivanov04} for a special case with equal strengthes of the rung coupling and
diagonal couplings. It was shown that the long-range ferrimagnetic state may
be destroyed and a singlet state appears at a large value of the frustration
parameter. However, it is not clear how the diagonal frustrations compete
with the exchanges along the legs and whether an exotic intermediate phase
exists for arbitrary rung coupling. Furthermore, an analytical analysis
which works in both regimes with weak and strong frustrations is still lack.
The general analytical method based on the spin wave theory is not capable
to deal with the regime with strong frustration where the long-range order
is destroyed. In order to provide a general understanding to frustration
effect in the ferrimagnetic ladder system, in this work we carry out
analytical investigation on the extended antiferromagnetic mixed-spin
ladders in the limit of strong rung couplings. When the rung coupling is
dominated, it is convenient and natural to map the original model onto an
effective spin-1/2 exchange model. Based on the effective model, we show
that the ground-state properties for both the frustrated and non-frustrated
ladders can be understood within the same framework and the phase transition
from a ferrimagnetic state to a critical spin-liquid state is also
discussed. In our theory, we treat the weak and strong frustrations on same
footing. For generality, we consider the most general frustrated ladder
model with arbitrary strengthes of exchanges of spins along the two legs as
well as the diagonal exchanges. Our model can interpolate between a variety
of systems, exhibiting remarkably rich ground state behavior with both
ordered and disordered phases.

\section{Model and the effective Hamiltonian}

As shown in Fig.1, the extended frustrated mixed-spin ladder is made of two
types of spins with magnitude $s_{_{i,1}}=1/2$ and $S_{i,2}=1$ located on
the rungs of the ladder. The corresponding Hamiltonian of the mixed-spin
ladder takes the form
\begin{equation}
H=H_0+H^{\prime }  \label{Hladder}
\end{equation}
with
\begin{equation}
H_0=\sum_{i=1}^NH_i=J_{\perp }\sum_{i=1}^N\widehat{s}_{i,1}\cdot \widehat{S}%
_{i,2}
\end{equation}
and
\begin{equation}
H^{\prime }=\sum_{i=1}^NH_{i,i+1}
\end{equation}
where
\begin{equation}
H_{i,i+1}=H_{i,i+1}^{leg_1}+H_{i,i+1}^{leg_2}+H_{i,i+1}^{d_1}+H_{i,i+1}^{d_2}
\end{equation}
and
\begin{eqnarray}
H_{i,i+1}^{leg_1} &=&J_1\widehat{s}_{i,1}\cdot \widehat{s}_{i+1,1}  \nonumber
\\
H_{i,i+1}^{leg_2} &=&J_2\widehat{S}_{i,2}\cdot \widehat{S}_{i+1,2}  \nonumber
\\
H_{i,i+1}^{d_1} &=&J_{\times }\widehat{s}_{i,1}\cdot \widehat{S}_{i+1,2}
\nonumber \\
H_{i,i+1}^{d_2} &=&J_{\times }^{\prime }\widehat{S}_{i,2}\cdot \widehat{s}%
_{i+1,1}.
\end{eqnarray}
Here $N$ is the number of rungs, and $\widehat{s}_{i,1}$ and $\widehat{S}%
_{i,2}$ represent the spin-$1/2$ and spin-1 operators respectively. For the
antiferromagnetic ladder, all the coupling parameters $J_{\perp
},J_1,J_2,J_{\times },J_{\times }^{\prime }>0$. The intrachain couplings
along the up and down legs are denoted by $J_1$ and $J_2$ respectively. The
interchain coupling across the rungs is $J_{\perp }$ and the diagonal
exchanges between the rungs are $J_{\times }$ and $J_{\times }^{\prime }$.
The summation is over the length of the chains and the periodic boundary is
assumed. For such a ladder model, the frustration comes from the competition
between the intrachain couplings ($J_1$ and $J_2)$ and the diagonal
couplings ($J_{\times }$ and $J_{\times }^{\prime })$. The model of (\ref
{Hladder}) covers a variety of known models. The model with $J_1=J_2=0$
corresponds to a bipartite mixed-spin ladder model whereas the one with $%
J_{\times }=J_{\times }^{\prime }=0$ is equivalent to the railroad ladder
model; For $J_1=J_2$ and $J_{\times }=J_{\times }^{\prime }$, our model
reduces to the ladder model investigated in Ref.\cite{Ivanov04}, where only
the case with $J_{\perp}=J_{\times }$ was considered; For $J_1=J_2$ and $%
J_{\times }^{\prime }$=0, the model is the dimerized zigzag mixed-spin ladder%
\cite{Ivanov,Pati}.

In this work, we will focus on the strong coupling limit with $J_{\perp }\gg
J_1,J_2,J_{\times },$ $J_{\times }^{\prime }$. In this limit, the
interactions between the neighboring rungs can be treated as perturbations
of the system of uncoupled rungs. It is instructive to start by considering
the two-site problem on an isolated rung with the Hamiltonian given by
\[
H_i=J_{\perp }\widehat{s}_{i,1}\cdot \widehat{S}_{i,2}.
\]
Eigenstates of the local Hamiltonian on a rung can be classified according
to the value of the total rung spin $S_i$. The two spins with magnitude $%
s_{i,1}=1/2$ and $S_{i,2}=1$ can combine into $S_i=1/2$ and $3/2$. It is
easy to get the eigenenergy $E_{1/2}=-$ $J_{\perp }$ and $E_{3/2}=J_{\perp
}/2$. The corresponding eigenstates are
\begin{eqnarray*}
\left| D_{\frac 12}\right\rangle _i &=&\frac 1{\sqrt{3}}\left[ -\left| \frac
12,0\right\rangle _i+\sqrt{2}\left| -\frac 12,1\right\rangle _i\right] , \\
\left| D_{-\frac 12}\right\rangle _i &=&\frac 1{\sqrt{3}}\left[ \left|
-\frac 12,0\right\rangle _i-\sqrt{2}\left| \frac 12,-1\right\rangle _i\right]
\end{eqnarray*}
for the doublet and
\begin{eqnarray*}
\left| Q_{\frac 32}\right\rangle _i &=&\left| \frac 12,1\right\rangle _i, \\
\left| Q_{\frac 12}\right\rangle _i &=&\frac 1{\sqrt{3}}\left[ \sqrt{2}%
\left| +\frac 12,0\right\rangle _i+\left| -\frac 12,1\right\rangle _i\right]
, \\
\left| Q_{-\frac 12}\right\rangle _i &=&\frac 1{\sqrt{3}}\left[ \sqrt{2}%
\left| -\frac 12,0\right\rangle _i+\left| \frac 12,-1\right\rangle _i\right]
, \\
\left| Q_{-\frac 32}\right\rangle _i &=&\left| -\frac 12,-1\right\rangle _i
\end{eqnarray*}
for the quartet, where $\left| s_z,S_z\right\rangle _i=\left|
s_z\right\rangle _{i,1}\otimes \left| S_z\right\rangle _{i.2}$. It is
obvious that the ground state of a rung is a doublet and the excited state
is a quartet with an excitation energy gap $3J_{\perp }/2.$ Therefore in the
strong rung-coupling limit spins on each rung of the ladder favor forming a
doublet. Since each rung can be either in the doubly degenerate state $%
\left| D_{\frac 12}\right\rangle $ or $\left| D_{-\frac 12}\right\rangle ,$
the ground state of the zero-order Hamiltonian $H_0$ is $2^N-$foldly
degenerate. When we go beyond the two-site problem, we need consider the
inter-rung couplings. In general, the inter-rung exchanges will lift the
degeneracy of the zero-order ground state and leads to an effective
Hamiltonian acting on the groundstate Hilbert space of $H_0$.

We then derive the effective Hamiltonian of the original ladder model in the
truncated Hilbert space composed of product of rung doublets by using
perturbation method. Similar schemes have been applied to study uniform spin
ladders in the strong-coupling limit\cite{Reigrotzki,Mila}. To the first
order and up to a constant of $-NJ_{\perp }$, the effective Hamiltonian can
be represented as
\begin{equation}
H_{eff}^{(1)}=\sum_i\left\langle \mu _{i,i+1}\right| H_{i,i+1}\left|
\upsilon _{i,i+1}\right\rangle \left| \mu _{i,i+1}\right\rangle \left\langle
\upsilon _{i,i+1}\right|  \label{eff1}
\end{equation}
where $\{\left| \mu _{i,i+1}\right\rangle ,\left| \upsilon
_{i,i+1}\right\rangle =\left| D_{\pm \frac 12}\right\rangle _i\otimes \left|
D_{\pm \frac 12}\right\rangle _{i+1}\}$ are four-foldly degenerate. It is
convenient to introduce pseudo-spin-$1/2$ operators $\widehat{\tau }_i$
which act on the states $\left| D_{1/2}\right\rangle _i$ and $\left|
D_{-1/2}\right\rangle _i$ and are defined as
\begin{eqnarray*}
\widehat{\tau }_i^z\left| D_{\pm \frac 12}\right\rangle _i &=&\pm \frac
12\left| D_{\pm \frac 12}\right\rangle _i, \\
\widehat{\tau }_i^{+}\left| D_{-\frac 12}\right\rangle _i &=&\left|
D_{+\frac 12}\right\rangle _i, \\
\widehat{\tau }_i^{-}\left| D_{+\frac 12}\right\rangle _i &=&\left|
D_{-\frac 12}\right\rangle _i, \\
\widehat{\tau }_i^{+}\left| D_{+\frac 12}\right\rangle _i &=&\widehat{\tau }%
_i^{-}\left| D_{-\frac 12}\right\rangle _i=0.
\end{eqnarray*}
With the above notation, we can identify the following relations
\begin{eqnarray*}
\widehat{\tau }_i^z &=&\frac 12\left| D_{+\frac 12}\right\rangle _{i\left.
{}\right. i}\left\langle D_{+\frac 12}\right| -\frac 12\left| D_{-\frac
12}\right\rangle _{i\left. {}\right. i}\left\langle D_{-\frac 12}\right| , \\
\widehat{\tau }_i^{+} &=&\left| D_{+\frac 12}\right\rangle _{i\left.
{}\right. i}\left\langle D_{-\frac 12}\right| , \\
\widehat{\tau }_i^{-} &=&\left| D_{-\frac 12}\right\rangle _{i\left.
{}\right. i}\left\langle D_{+\frac 12}\right| .
\end{eqnarray*}
Therefore we can rewrite terms of $\left| \mu _{i,i+1}\right\rangle
\left\langle \upsilon _{i,i+1}\right| $ in terms of the pseudo-spin
operators. For example, we have
\[
\left| \mu _{i,i+1}\right\rangle \left\langle \upsilon _{i,i+1}\right| =%
\widehat{\tau }_i^{+}\widehat{\tau }_{i+1}^{-}
\]
for $\left| \mu _{i,i+1}\right\rangle =$ $\left| D_{+\frac 12}\right\rangle
_i\left| D_{-\frac 12}\right\rangle _{i+1}$ and $\left| \nu
_{i,i+1}\right\rangle =\left| D_{-\frac 12}\right\rangle _i\left| D_{+\frac
12}\right\rangle _{i+1}$ and the corresponding coefficient is given by
\[
\left\langle \mu _{i,i+1}\right| H_{ij}\left| \nu _{i,i+1}\right\rangle
=-\frac 29(J_{\times }+J_{\times }^{\prime })+\frac 1{18}J_1+\frac 89J_2.
\]
After some algebras, we can rewrite the effective Hamiltonian (\ref{eff1})
as the following form
\begin{equation}
H_{eff}^{(1)}/J_{\perp }=\sum_iJ_{eff}^{(1)}\left[ \frac 12\left( \widehat{%
\tau }_i^{+}\widehat{\tau }_{i+1}^{-}+\widehat{\tau }_i^{-}\widehat{\tau }%
_{i+1}^{+}\right) +\widehat{\tau }_i^z\widehat{\tau }_{i+1}^z\right]
\label{Heff}
\end{equation}
with the effective parameter given by
\begin{equation}
J_{eff}^{(1)}=-\frac 49\left( \frac{J_{\times }}{J_{\perp }}+\frac{J_{\times
}^{\prime }}{J_{\perp }}\right) +\frac 19\frac{J_1}{J_{\perp }}+\frac{16}9%
\frac{J_2}{J_{\perp }}.  \label{Jeff}
\end{equation}

Next we carry out calculation of the second order perturbation which is
expected to give correction to the effective coupling parameter and the
ground energy. In the first order calculation, the higher excited states
(quartets) on the isolated rungs do not play role. However, such rung
quartets give contribution to the higher order correction. The second order
correction can be described by
\begin{eqnarray}
H_{eff}^{\left( 2\right) } &=&\sum_{i,m\neq 0}\frac{\left\langle \mu
_{i,i+1}\right| H_{i,i+1}\left| m\right\rangle \left\langle m\right|
H_{i,i+1}\left| \upsilon _{i,i+1}\right\rangle }{E_0-E_m}  \nonumber \\
&&\times \left| \mu _{i,i+1}\right\rangle \left\langle \upsilon
_{i,i+1}\right|
\end{eqnarray}
where $E_0=2E_{1/2}=-2J_{\perp }$ and $\left| m\right\rangle $ are the
intermediate states with at least one quartet residing in the neighboring $i$
th and $\left( i+1\right) $th rungs. Therefore $E_m=E_{1/2}+E_{3/2}=-J_{%
\perp }/2$ if $\left| m\right\rangle =$ $\{\left| D_\alpha \right\rangle
_i\left| Q_\beta \right\rangle _{i+1}$ or $\left| Q_\beta \right\rangle
_i\left| D_\alpha \right\rangle _{i+1}\},$ and $E_m=2E_{3/2}=J_{\perp }$ if $%
\left| m\right\rangle =\left| Q_\beta \right\rangle _i\left| Q_{\beta
^{\prime }}\right\rangle _{i+1}$ where $\alpha =-1/2,1/2,$ and $\beta ,\beta
^{\prime }=-3/2,-1/2,1/2,3/2.$ After tedious but straightforward
calculation, we can represent $H_{eff}^{\left( 2\right) }$ in terms of the
pseudo-spin operators
\begin{equation}
H_{eff}^{(2)}/J_{\perp }=\sum_i\left[ J_{eff}^{(2)}\widehat{\tau }_i\cdot
\widehat{\tau }_{i+1}+c^{\prime }\right] ,
\end{equation}
where
\begin{eqnarray}
J_{eff}^{(2)} &=&-\frac 4{243}\left( \frac{J_1}{J_{\perp }}+4\frac{J_2}{%
J_{\perp }}-4\frac{J_{\times }}{J_{\perp }}-\frac{J_{\times }^{^{\prime }}}{%
J_{\perp }}\right) ^2  \nonumber \\
&&-\frac 4{243}\left( \frac{J_1}{J_{\perp }}+4\frac{J_2}{J_{\perp }}-\frac{%
J_{\times }}{J_{\perp }}-4\frac{J_{\times }^{^{\prime }}}{J_{\perp }}\right)
^2  \nonumber \\
&&+\frac 8{243}\left( \frac{J_1}{J_{\perp }}+\frac{J_2}{J_{\perp }}-\frac{%
J_{\times }}{J_{\perp }}-\frac{J_{\times }^{^{\prime }}}{J_{\perp }}\right)
^2  \label{Jeff2}
\end{eqnarray}
and
\begin{eqnarray}
c^{\prime } &=&-\frac 4{81}\left( \frac{J_1}{J_{\perp }}+\frac{J_2}{J_{\perp
}}-\frac{J_{\times }}{J_{\perp }}-\frac{J_{\times }^{^{\prime }}}{J_{\perp }}%
\right) ^2  \nonumber \\
&&-\frac 1{81}\left( \frac{J_1}{J_{\perp }}+4\frac{J_2}{J_{\perp }}-4\frac{%
J_{\times }}{J_{\perp }}-\frac{J_{\times }^{^{\prime }}}{J_{\perp }}\right)
^2  \nonumber \\
&&-\frac 1{81}\left( \frac{J_1}{J_{\perp }}+4\frac{J_2}{J_{\perp }}-\frac{%
J_{\times }}{J_{\perp }}-4\frac{J_{\times }^{^{\prime }}}{J_{\perp }}\right)
^2.  \label{c'}
\end{eqnarray}
The second order correction can also produce a term due to the three-site
process which is described by
\begin{eqnarray*}
&&\left\langle \mu _{i,i+1,i+2}\right| H_{eff}^{\prime \left( 2\right)
}\left| \upsilon _{i,i+1,i+2}\right\rangle  \\
&=&\sum_{i,m\neq 0}\left[ \frac{\left\langle \mu _{i,i+1,i+2}\right|
H_{i,i+1}\left| m\right\rangle \left\langle m\right| H_{i+1,i+2}\left|
\upsilon _{i,i+1,i+2}\right\rangle }{E_0-E_m}\right.  \\
&&+\left. \frac{\left\langle \mu _{i,i+1,i+2}\right| H_{i+1,i+2}\left|
m\right\rangle \left\langle m\right| H_{i,i+1}\left| \upsilon
_{i,i+1,i+2}\right\rangle }{E_0-E_m}\right]
\end{eqnarray*}
where $E_0=3E_{1/2}=-3J_{\perp }$ and $\left| m\right\rangle $ are the
intermediate states with a quartet residing in the $\left( i+1\right) $th
rung. In terms of the pseudo-spin operators, the Hamiltonian $%
H_{eff}^{\prime \left( 2\right) }$ is finally simplified to
\begin{equation}
H_{eff}^{\prime \left( 2\right) }/J_{\perp }=\sum_{i=1}^NJ_{eff}^{\prime }%
\widehat{\tau }_i\cdot \widehat{\tau }_{i+2}
\end{equation}
with
\begin{eqnarray}
J_{eff}^{\prime } &=&-\frac 8{243}\left( \frac{J_1}{J_{\perp }}+4\frac{J_2}{%
J_{\perp }}-\frac{J_{\times }}{J_{\perp }}-4\frac{J_{\times }^{^{\prime }}}{%
J_{\perp }}\right) \times   \nonumber \\
&&\left( \frac{J_1}{J_{\perp }}+4\frac{J_2}{J_{\perp }}-4\frac{J_{\times }}{%
J_{\perp }}-\frac{J_{\times }^{^{\prime }}}{J_{\perp }}\right) .
\end{eqnarray}

Therefore, up to the second order, the effective Hamiltonian of the original
ladder can be written as
\begin{equation}
H_{eff}/J_{\perp }=\sum_{i=1}^N\left( J_{eff}\widehat{\tau }_i\cdot \widehat{%
\tau }_{i+1}+J_{eff}^{\prime }\widehat{\tau }_i\cdot \widehat{\tau }%
_{i+2}+c\right) ,  \label{H_eff}
\end{equation}
where $J_{eff}=J_{eff}^{(1)}+J_{eff}^{(2)}$ and $c=-1+c^{\prime }$. The
effective Hamiltonian (\ref{H_eff}) describes a spin-$1/2$ Heisenberg chain
with additional next-nearest-neighbor (NNN) exchanges.

\section{Phase diagram and discussion}

It is convenient to determine the phase diagram of the original
ladder model from the effective Hamiltonian (\ref{H_eff}) since it
has a much simpler form and was widely studied. Since the terms
with NNN exchanges come from the second-order perturbation, in
general we always have$\left| J_{eff}^{\prime }\right| \ll \left|
J_{eff}\right| $.  A small NNN ferromagnetic interaction is an
irrelevant perturbation to the spin chain
model\cite{Chen2,Haldane}. Therefore we may approximate to
determine the phase diagram by omitting the term of
$J_{eff}^{\prime }.$ The effective Hamiltonian (\ref{H_eff}) with
$J_{eff}^{\prime }=0$ describes either a ferromagnetic or an
antiferromagnetic spin-$1/2$ Heisenberg chain depending on the
sign of the effective parameter. Based on the effective
Hamiltonian,
one is ready to get the ground energy of the mixed-spin ladder (\ref{Hladder}%
) in the strong coupling limit:
\begin{equation}
E_g/NJ_{\perp }=\varepsilon _H+\delta \varepsilon ^{\prime }+c,
\end{equation}
where $\varepsilon _H$ is the ground energy per site of an isotropic spin-$%
1/2$ Heisenberg chain and $\delta \varepsilon ^{\prime }$ is the energy
correction due to the NNN exchanges which reads
\[
\delta \varepsilon ^{\prime }=\left\langle GS\right| H_{eff}^{\prime \left(
2\right) }\left| GS\right\rangle /NJ_{\perp }
\]
with $\left| GS\right\rangle $ representing the ground state of Heisenberg
chain. From the corresponding results of the well-known Heisenberg chain\cite
{Takahashi,Mattis}, we have
\[
\varepsilon _H/J_{eff}=\left\{
\begin{array}{lll}
1/4, &  & \text{for }J_{eff}<0 \\
1/4-\ln 2, &  & \text{for }J_{eff}>0
\end{array}
\right. .
\]
When the system is in the ferromagnetic state, it is ready to obtain
\[
\delta \varepsilon ^{\prime }=J_{eff}^{\prime }/4.
\]
When the system is in the antiferromagnetic phase, no an analytical result
is available.

From eq.(\ref{Jeff}), we see that the diagonal exchanges tend to produce an
effective ferromagnetic coupling whereas the exchanges along the legs lead
to an effective antiferromagnetic coupling. Both the ferromagnetic and
antiferromagnetic spin-$1/2$ Heisenberg chains are exactly solved by the
Bethe-ansatz method\cite{Takahashi,Mattis}, but they have very different
ground-state properties. The ferromagnetic ground state consists of all
spins parallel and thus has long-range order, whereas the antiferromagnetic
ground state is a spin singlet which can be described by the critical
spin-liquid phase. If $J_{eff}<0,$ the effective model is a ferromagnetic
chain which exhibits gapless excitations and ferromagnetic ground state.
Equivalently, within the first order approximation the original mixed-spin
ladder has ferrimagnetic ground state if $J_{\times }+J_{\times }^{\prime
}>J_1/4+4J_2.$ On the other hand, the effective model is an
antiferromagnetic chain if $J_{eff}>0$. Correspondingly, the ground state of
the original mixed-spin ladder is a critical spin liquid if $J_{\times
}+J_{\times }^{\prime }<J_1/4+4J_2$.
\begin{figure}[tbp]
\includegraphics[width=3.4in]{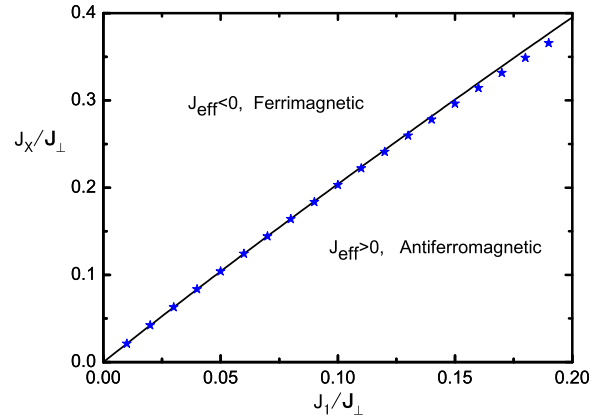}
\caption{Phase diagram of the mixed-spin ladder with $J_1=J_2$ and $%
J_{\times}=J^{\prime}_{\times}$ in the strong coupling limit. The solid line
is the phase boundary determined by $J_{eff}=0$ whereas the separated dots
denoted by stars are obtained from the exact diagonalization of the spin
ladder model.}
\label{fig2}
\end{figure}

As we have mentioned that our mixed-spin ladder model includes a series of
models as its special cases, we may understand the ground-state properties
of these models by using our derived effective Hamiltonian (\ref{H_eff}).
When $J_1=J_2=0,$ our model reduces to a bipartite mixed-spin ladder model.
The effective model (\ref{H_eff}) is a ferromagnetic Heisenberg chain
because we have $J_{eff}<0.$ Therefore we can conclude that the total spin
of the ground state is $N/2.$ When $J_{\times }=J_{\times }^{\prime }=0,$ we
get the railroad ladder model composed of coupled spin-$1/2$ and spin-$1$
chains. It is ready to conclude that its ground state is a singlet because
the effective model (\ref{H_eff}) is an antiferromagnetic Heisenberg chain
with $J_{eff}>0.$ For the frustrated spin ladder with $J_1=J_2$ and $%
J_{\times }=J_{\times }^{\prime }$, it is easy to see from the effective
model (\ref{H_eff}) that there exists a quantum phase transition arising
from the competition of $J_1$ and $J_{\times }.$ The transition point is
determined by $J_{eff}=0$ with
\begin{equation}
J_{eff}=-\frac 89\frac{J_{\times }}{J_{\perp }}+\frac{17}9\frac{J_1}{%
J_{\perp }}-\frac{56}{81}\left( \frac{J_1}{J_{\perp }}-\frac{J_{\times }}{%
J_{\perp }}\right) ^2 .
\end{equation}
From the solution of the above equation, we can get the phase diagram of the
mixed spin ladder as shown in Fig.2. Above the phase boundary the original
ladder is in a ferrimagnetic phase corresponding to $J_{eff}<0$, whereas the
system is in an antiferromagnetic phase with $J_{eff}>0$ below the phase
boundary.

In order to check how good is our theory based on the perturbation method
and the validation of the phase boundary determined by $J_{eff}=0$, we study
the original spin ladder model by exact diagonalization method and determine
its phase diagram numerically. The phase boundary determined by the
effective spin chain model is not sensitive to the size of the system
(Actually the phase boundary is completely independent of the size when we
omit the term of NNN interaction). Therefore for the purpose of verification
of our analytical result, it is enough for us to consider a $2\times4$-size
ladder which can be diagonalized by the exact diagonalization method. The
phase boundary can be determined simply by the ground state degeneracy of
the system. In comparison with phase boundary obtained by omitting the term
of the NNN exchanges, we find that they agree very well in the strong
coupling limit and as expected, the analytical result begins to deviate the
exact numerical result when $J_1/J_{\perp}$ increases.

Finally, we address a special point with $J_1=J_2=J_{\times
}=J_{\times }^{\prime }$ where the total spin on a rung is a good
quantum number. From eq.(\ref{Jeff}), it is obvious that the model
is effectively described by an antiferromagnetic spin chain with
an effective coupling constant of $J_1$ and the high-order
corrections vanish due to $J^{(2)}_{eff}=J'_{eff}=0$. Actually
this is an exact conclusion for this special case. This is clear
if we rewrite the original Hamiltonian as
\begin{equation}
H=\sum_{i=1}^N\left[ \frac{J_{\perp }}2\widehat{\mathbf{S}}_i^2+J_1\widehat{%
\mathbf{S}}_i\cdot \widehat{\mathbf{S}}_{i+1}-\frac{11J_{\perp }}8\right] ,
\end{equation}
where $\widehat{\mathbf{S}}_i^2=S_i(S_i+1)$ with $S_i=1/2$ or $3/2$ and $%
\widehat{\mathbf{S}}_i$ is defined as $\widehat{\mathbf{S}}_i\equiv $ $%
\widehat{s}_{i,1}+\widehat{S}_{i,2}$ denoting the total spin on the $i$th
rung. In the strong coupling limit $J_{\perp }\gg J_1$, $J_{\perp }$ forces $%
{S}_i$ to take the value of $1/2$ and thus the ground-state properties of $H$
is described by an effective spin-$1/2$ antiferromagnetic Heisenberg chain.
From eqs.(\ref{Jeff2}) and (\ref{c'}), we see that $J_{eff}^{(2)}=c^{\prime
}=0$ when $J_1=J_2=J_{\times }=J_{\times }^{\prime }$. This implies that the
higher order correction also conforms to the exact result in this special
case.

\section{Conclusions}

In conclusion, we have studied the ground-state properties of a
generalized mixed-spin ladder with diagonal exchanges in the limit
of strong rung couplings. By mapping it to an effective spin-$1/2$
Heisenberg chain with additional NNN exchanges, we find that the
diagonal exchanges lead to the ferromagnetic effective coupling
whereas the exchanges along the legs produces the
antiferromagnetic effective coupling. The ground state of the
effective Hamiltonian is either ferromagnetic or antiferromagnetic
depending on the competition between these two opposite processes.
With the help of the effective Hamiltonian and omitting the
additional small NNN exchanges, it is straightforward to
analytically determine the transition point from the ferrimagnetic
phase to the critical spin liquid phase. The phase boundary is
found to agree with the result obtained by exact numerical
diagonaliztion of the original spin ladder model. Our results show
that the strong coupling approach provides a simple and unifying
way to exhibit the rich physics of the mixed-spin ladders.

\begin{acknowledgments}
This work is supported by NSF of China under Grant No. 10574150
and programs of Chinese Academy of Sciences.
\end{acknowledgments}


\begin{references}
\bibitem{Kahn}  O. Kahn, Y. Pei, and Y. Journaux, in {\it Inorganic Materials
}, edited by D. W. Bruce and D. O'Hare (John Wiley \& Sons Ltd.,
New York, 1992); O. Kahn, Struct. Bonding (Berlin) {\bf 68}, 89
(1987).

\bibitem{Hagiwara} M. Hagiwara, K. Minami, Y. Narumi, K. Tatani, and K. Kindo, J.
Phys. Soc. Jpn. {\bf 67}, 2209 (1998); {\bf 68}, 2214 (1999); N.
Fujiwara and M. Hagiwara, Solid State Commun. {\bf 113}, 443
(2000).

\bibitem{Verdaguer}
M. Verdaguer, A. Gleizes, J. P. Renard, and J. Selden, Phys. Rev.
B {\bf 29}, 5144 (1984); Y. Hosokoshi, Y. Nakazawa, K. Indue, K.
Takizawa, H. Nakano, M. Takahashi, and M. Goto, Phys. Rev. B {\bf
60}, 12924 (1999).

\bibitem{Brehmer}  S. Brehmer, H.-J. Mikeska, and S. Yamamoto, J. Phys.:
Condens. Matter {\bf 9}, 3921 (1997).

\bibitem{Pati}  S. K. Pati, S. Ramasesha, and D. Sen, Phys. Rev. B {\bf 55},
8894 (1997) ; J. Phys.: Condens. Matter {\bf 9}, 8707 (1997).

\bibitem{Kolezhuk}  A. K. Kolezhuk, H.-J. Mikeska, and S. Yamamoto, Phys.
Rev. B {\bf 55}, R3336 (1997).

\bibitem{Tian}  G.-S. Tian, Phys. Rev. B {\bf 56}, 5355 (1997).

\bibitem{Wu}  C. Wu, B. Chen, X. Dai, Y. Yu, and Z.-B. Su, Phys. Rev. B {\bf
60}, 1057 (1999).

\bibitem{Kolezhuk2} A. K. Kolezhuk and H.-J. Mikeska, Eur. Phys.  J. B {\bf 5}, 543
(1998).

\bibitem{Langari2000}  A. Langari, M. Abolfath and M. A. Martin-Delgado,
Phys. Rev. B, {\bf 61}, 343 (2000).

\bibitem{Langari}  A. Langari and M. A. Martin-Delgado, Phys. Rev. B, {\bf 63%
}, 054432 (2001).

\bibitem{Trumper}  E. Trumper and C. Gazza, Phys. Rev. B, {\bf 64}, 134408
(2001).

\bibitem{Aristov}  D. N. Aristov and M. N. Kiselev, Phys. Rev. B, {\bf 70},
224402 (2004).

\bibitem{Ivanov}  N. B. Ivanov, J. Richter and U. Schollwock, Phys. Rev. B,
{\bf 58}, 14456 (1998).

\bibitem{Ivanov04}  N. B. Ivanov and J. Richter, Phys. Rev. B, {\bf 69},
214420 (2004).

\bibitem{Chen}  S. Chen and B. Han, Eur. Phys. J. B {\bf 31}, 63,
(2003); S. Chen and H. Buttner, Eur. Phys. J. B {\bf 29}, 15,
(2002).

\bibitem{Chen2} S. Chen, H. Buttner and J. Voit, Phys. Rev. B {\bf 67},
054412, (2003); Phys. Rev. Lett. {\bf 87}, 087205, (2001).

\bibitem{Lecheminant} See, for example, the review article of P. Lecheminant in the book
``{\it Frustrated spin systems}", edited by H. T. Diep, World
Scientific, Singapore (2003), and references therein.

\bibitem{Reigrotzki}  M. Reigrotzki, H. Tsunetsugu and T. M. Rice, J. Phys.:
Condens. Matter {\bf 6,} 9235 (1994).

\bibitem{Mila}  F. Mila, Eur. Phys. J. B. {\bf 6}, 201 (1998).

\bibitem{Haldane} F. D. M. Haldane, Phys. Rev. B
{\bf 25}, 4925 (1982).

\bibitem{Takahashi} M. Takahashi, \textsl{Thermodynamic of One-Dimensional Solvable
Models.} (Cambridge University Press, Cambridge 1999).

\bibitem{Mattis} D. C. Mattis, \textsl{The Many-body Problem: An Encyclopedia of Exactly
Solved Models in One Dimension} (World Scientific, Singapore
1993).


\end{references}
\end{document}